OITS-598

# Remarks on a Technique of Measuring CP phase $\alpha$

N.G. Deshpande[1], Xiao-Gang He[2] and Sechul Oh[1]

[1]*Institute of Theoretical Science*

*University of Oregon*

*Eugene, OR 97403-5203, USA*

*and*

[2]*School of Physics, University of Melbourne*

*Parkville, Vic. 3052, Australia*

(February, 1996)

## Abstract

We present a method to measure the CKM phase $\alpha$ and the tree and penguin (strong and electroweak) amplitudes in $B \to \pi\pi$ decays, based on isospin consideration and the weak assumption that all tree amplitudes have a common strong phase and all penguin amplitudes have a different common phase. The method needs only the time-independent measurements of the relevant decay rates in $B \to \pi\pi$. We also propose a method to experimentally examine the validity of the assumption that all penguin amplitudes have the same strong phases, and to extract detailed informations about the hadronic matrix elements.

The Standard Model of three generations with the source for CP violation arising from the phases in the Cabibbo-Kobayashi-Maskawa (CKM) matrix is so far consistent with the experiment [1]. An important way of verifying the CKM model is to measure the three angles $\alpha \equiv \mathrm{Arg}(-V_{td}V_{tb}^{*}/V_{ub}^{*}V_{ud})$, $\beta \equiv \mathrm{Arg}(-V_{cd}V_{cb}^{*}/V_{tb}^{*}V_{td})$, and $\gamma \equiv \mathrm{Arg}(-V_{ud}V_{ub}^{*}/V_{cb}^{*}V_{cd})$ of the unitarity triangle of the CKM matrix independently in experiment and to check whether the sum of these three angles is equal to $180^o$, as it should be in the model. $B$ meson decays provide a fertile ground to carry out such a test [2,3]. One class of methods to measure the CKM phases involve the measurements of CP asymmetries in time evolution of $B^0$ decays into CP eigenstates [3]. Since most decay processes get contributions from both tree and loop (penguin) effects, in order to measure the CKM phases without hadronic uncertainties, in many cases one needs additional information such as using relations based on isospin or flavor SU(3) symmetries [4–10]. The phase $\beta$ can be determined unambiguously by measuring CP asymmetry in time evolution of $B \to \Psi K_S$ [3]. The phase $\gamma$ can be measured in $B_S^0 \to \rho K_S^0$, $B^- \to DK^-$ [11], or by the methods using information from $B \to \pi\pi$ and $B \to \pi(\eta)K$ based on flavor SU(3) symmetries [7].

The extraction of $\alpha$ involves the study of CP asymmetry in $B \to \pi\pi$ or $B \to \rho\pi$ [4,5]. If the penguin contributions are neglected, this extraction is straightforward. However, if penguin diagrams make a significant contribution, then the interpretations of the results become complicated. Gronau and London presented a method using an isospin analysis of $B \to \pi\pi$ and $\bar{B} \to \pi\pi$ to remove the (strong) penguin contamination [4]. It was shown that the inclusion of the electroweak penguin contributions makes only a small error in $\alpha$ determination [8]. So their method provides a relatively clean way to determine $\alpha$, even though the method still needs the time-dependent measurement of CP asymmetry.

In a recent letter [9] interesting method to approximately measure the phase $\alpha$ without using time-dependent asymmetry in $B(\bar{B}) \to \pi\pi$ has been proposed by Hamzaoui and Xing (HX). This method is based on isospin consideration, and might have advantages, since it needs only the time-independent measurements of the relevant decay rates, which can be carried out at either $B$ factories or high-luminosity hadron machines and it can give an estimation for $\alpha$ from the data of $B \to \pi\pi$, before an accurate measurement of $\alpha$ is available from an asymmetric collider. However, in a recent paper [12] we showed that this method actually fails because the neglected electroweak penguin effects turn out to be large and invalidate this method. In this letter we improve the HX approach by including the electroweak penguin effects and show that under certain assumptions one can determine the phase $\alpha$ as well as the tree and penguin amplitudes using $B \to \pi\pi$ decays. We also show that $B \to \pi\pi$ decays can be used to carry out detailed study of relevant hadronic matrix elements and strong phases.

Based on isospin consideration we parameterize the decay amplitudes for $B \to \pi\pi$ as follows:

$$\begin{aligned} A_{+-} &\equiv <\pi^+\pi^-|H|B^0>= -Te^{i\gamma} - P_{+-}e^{i(\delta_{+-}-\beta)}, \\ A_{00} &\equiv <\pi^0\pi^0|H|B^0>= \frac{a}{\sqrt{2}}Te^{i\gamma} - \frac{1}{\sqrt{2}}P_{00}e^{i(\delta_{00}-\beta)}, \\ A_{+0} &\equiv <\pi^+\pi^0|H|B^+>= -\frac{1+a}{\sqrt{2}}Te^{i\gamma} - \frac{1}{\sqrt{2}}P_{+0}e^{i(\delta_{+0}-\beta)}; \end{aligned} \tag{1}$$

and

$$\begin{aligned} \bar{A}_{+-} &\equiv <\pi^+\pi^-|H|\bar{B}^0>= -Te^{-i\gamma} - P_{+-}e^{i(\delta_{+-}+\beta)}, \\ \bar{A}_{00} &\equiv <\pi^0\pi^0|H|\bar{B}^0>= \frac{a}{\sqrt{2}}Te^{-i\gamma} - \frac{1}{\sqrt{2}}P_{00}e^{i(\delta_{00}+\beta)}, \\ \bar{A}_{-0} &\equiv <\pi^-\pi^0|H|\bar{B}^->= -\frac{1+a}{\sqrt{2}}Te^{-i\gamma} - \frac{1}{\sqrt{2}}P_{+0}e^{i(\delta_{+0}+\beta)}, \end{aligned} \tag{2}$$

where $\bar{A}_{ij}$'s denote the CP-conjugate amplitudes of $A_{ij}$'s. The $T$ and $P_{ij}$'s are the tree and (both strong and electroweak) penguin amplitudes,respectively, and can be assumed real and positive here. The parameter $a$ which arises from color suppressed contribution in the tree amplitudes can be considered a complex parameter in general. $\delta_{ij}$ are the strong phases of the penguin amplitudes. The isospin relations,

$$A_{+0} + A_{00} = \frac{1}{\sqrt{2}}A_{+-}\ , \quad \bar{A}_{+0} + \bar{A}_{00} = \frac{1}{\sqrt{2}}\bar{A}_{+-}\ , \tag{3}$$

implies:

$$P_{+0}e^{i\delta_{+0}} + P_{00}e^{i\delta_{00}} = P_{+-}e^{i\delta_{+-}} \ . \tag{4}$$

Although we have given a general parameterization of the decay amplitudes, we shall make a weak assumption that $a$ is real and that $\delta_{+0} = \delta_{00} = \delta_{+-} \equiv \delta$. Thus we assume that all tree amplitudes have a common strong phase, and all penguin amplitudes have a different common phase. This is certainly true in factorization approximation, where tree contribution has no absorptive part, and penguin contribution has absorptive part from the c and u loop which give approximately equal strong phases. This assumption, of course, must be checked by experiments. We will discuss how to test this assumption in future experiments later.

Now we define the following measurables

$$\eta_1 \equiv \frac{|A_{+0}|^2 - |\bar{A}_{-0}|^2}{|A_{+-}|^2 - |\bar{A}_{+-}|^2}, \quad \eta_2 \equiv \frac{|A_{00}|^2 - |\bar{A}_{00}|^2}{|A_{+-}|^2 - |\bar{A}_{+-}|^2}. \tag{5}$$

Then from Eqs.(1) and (2), the parameter $a$ is given as

$$a = \eta_1 - \eta_2 - \frac{1}{2} \pm \sqrt{(\eta_1 - \eta_2 - \frac{1}{2})^2 - 2\eta_2}. \tag{6}$$

The tree and penguin amplitudes are expressed in terms of the measurables:

$$\begin{aligned}
T &= \sqrt{\{\frac{2\eta_1(\eta_2|A_{+-}|^2 - |A_{00}|^2)}{(1+a)^2(a^2-2\eta_2)} - \frac{2\eta_2(\eta_1|A_{+-}|^2 - |A_{+0}|^2)}{a^2[(1+a)^2 - 2\eta_1]}\}/[\frac{\eta_2}{a^2} - \frac{\eta_1}{(1+a)^2}]}, \\
P_{+-} &= \sqrt{\{\frac{\eta_2|A_{+-}|^2 - |A_{00}|^2}{a^2-2\eta_2} - \frac{\eta_1|A_{+-}|^2 - |A_{+0}|^2}{(1+a)^2 - 2\eta_1}\}/[\frac{\eta_2}{a^2} - \frac{\eta_1}{(1+a)^2}]}, \\
P_{00} &= -\frac{2\eta_2}{a}P_{+-}, \\
P_{+0} &= (1 + \frac{2\eta_2}{a})P_{+-}.
\end{aligned} \tag{7}$$

We obtain

$$\begin{aligned}
\cos(\alpha+\delta) &= \frac{T^2 + P_{+-}^2 - |A_{+-}|^2}{2TP_{+-}} \\
\cos(\alpha-\delta) &= \frac{T^2 + P_{+-}^2 - |\bar{A}_{+-}|^2}{2TP_{+-}}.
\end{aligned} \tag{8}$$

Here we have used the relation $\alpha = 180^0 - \beta - \gamma$. From Eq.(8), we can determine both $\alpha$ and $\delta$ with eight-fold ambiguity. Half of the eight different solutions arises from the two different values of $a$ and may be eliminated. From factorization calculation, it can be shown that $|\eta_1|$ is much smaller than $|\eta_2|$. If this turns out to be true experimentally, the two solutions in Eq.(6) are approximately, $-1$ and $-2\eta_2$. The solution $a \approx -1$ implies that the color suppressed tree level contribution is the same in magnitude as the unsuppressed one. This solution is unlikely to be true. However, we are still left with four-fold ambiguity for the determination of $\alpha$. We remark in passing that if the electroweak penguins are neglected, we recover equations in Ref. [9]. The quantities $|A_{ij}|^2 - |\bar{A}_{ij}|^2$ can be measured at symmetric colliders [13]. The measurements for $|A_{+-}|^2 - |\bar{A}_{+-}|^2$ and $|A_{00}|^2 - |\bar{A}_{00}|^2$ may be difficult. In a recent paper by two of us [14], it was shown that there are relationships between rate differences: $\Delta(\bar{B}^0 \to \pi^+\pi^-(\pi^0\pi^0)) = \Gamma(\bar{B}^0 \to \pi^+\pi^-(\pi^0\pi^0)) - \Gamma(B^0 \to \pi^+\pi^-(\pi^0\pi^0))$ and $\Delta(\bar{B}^0 \to \pi^+K^-(\pi^0\bar{K}^0)) = \Gamma(\bar{B}^0 \to \pi^+K^-(\pi^0\bar{K}^0)) - \Gamma(B^0 \to \pi^-K^+(\pi^0K^0))$, which follows from SU(3) symmetry. Using the relationship, one can carry out the easier measurement of $\Delta(\bar{B}^0 \to \pi^+K^-(\pi^0\bar{K}^0))$ instead of $\Delta(\bar{B}^0 \to \pi^+\pi^-(\pi^0\pi^0))$.

Eventually, after $\alpha$, $\beta$ and $\gamma$ are determined in other experiments, it is possible to test if our hypothesis was valid. We present a method to determine whether the strong phase shifts $\delta_{ij}$'s corresponding to each decay amplitudes $A_{ij}$'s in $B \to \pi\pi$ are really all the same, still assuming that the parameter $a$ is real. This method can determine the tree and penguin amplitudes in $B \to \pi\pi$ as well. We assume that in addition to the magnitudes of $A_{ij}$'s and $\bar{A}_{ij}$'s in $B \to \pi\pi$, the CKM phases $\beta$ and $\gamma$ are the "known" quantities from the experimental measurements, say, using the decay modes $B \to \Psi K_S$ and $B^- \to K^-D$, and $\alpha$ is "known" from the unitarity condition $\alpha + \beta + \gamma = 180^o$. Besides, we assume that the parameters $Im\lambda_{ij} \equiv Im(e^{-2i\beta}\bar{A}_{ij}/A_{ij})$ are determined by the time-dependent measurements of CP asymmetry in $B^0 \to \pi^+\pi^-$ and $\pi^0\pi^0$.

It is convenient to define rotated amplitudes:

$$\tilde{A}_{ij} \equiv e^{-i(\gamma+\phi)}A_{ij}, \quad \tilde{\bar{A}}_{ij} \equiv e^{+i(\gamma-\phi)}\bar{A}_{ij}. \tag{9}$$

Note that $\tilde{\bar{A}}_{+-}/\tilde{A}_{+-} = e^{i\theta}|\bar{A}_{+-}/A_{+-}|$, where $\theta$ can be determined by measuring $Im\lambda_{+-}$. The phase $\phi$ is the total phase in $A_{+-}e^{-i\gamma}$. It is clear that the amplitude $\tilde{A}_{+-}$ is real. We will use it as the orientation axis for all other amplitudes. Since $\theta$ is assumed to be known, we can construct the two triangles in complex plane as shown in Fig.1, using the isospin relations among the rotated amplitudes, where from the information of $Im\lambda_{00}$ we can remove the possibility of the different orientations of the triangles. The quantities $F$, $G$, $\bar{F}$, and $\bar{G}$ defined in Fig.1 can be described as

$$\begin{aligned}
F = Re\tilde{A}_{+0} &= -\frac{1+a}{\sqrt{2}}T\cos\phi + \frac{P_{+0}}{\sqrt{2}}\cos(\delta_{+0}+\alpha-\phi), \\
G = Im\tilde{A}_{+0} &= \frac{1+a}{\sqrt{2}}T\sin\phi + \frac{P_{+0}}{\sqrt{2}}\sin(\delta_{+0}+\alpha-\phi), \\
\bar{F} = Re\tilde{\bar{A}}_{-0} &= -\frac{1+a}{\sqrt{2}}T\cos\phi + \frac{P_{+0}}{\sqrt{2}}\cos(\delta_{+0}-\alpha-\phi), \\
\bar{G} = Im\tilde{\bar{A}}_{-0} &= \frac{1+a}{\sqrt{2}}T\sin\phi + \frac{P_{+0}}{\sqrt{2}}\sin(\delta_{+0}-\alpha-\phi). \qquad (10)
\end{aligned}$$

We note that the quantities $F$, $G$, $\bar{F}$, and $\bar{G}$ are completely determined from Fig.1. From the figure, we can easily obtain the followings:

$$\begin{aligned}
\tan\phi &= -\frac{G+\bar{G}+(F-\bar{F})\cot\alpha}{F+\bar{F}-(G-\bar{G})\cot\alpha}, \\
(1+a)T &= -\frac{1}{\sqrt{2}\cos\phi}[F+\bar{F}-(G-\bar{G})\cot\alpha], \\
P_{+0} &= \frac{1}{\sqrt{2}\sin\alpha}\sqrt{(F-\bar{F})^2+(G-\bar{G})^2}, \\
\tan(\delta_{+0}-\phi) &= -\frac{\bar{F}-F}{\bar{G}-G}. \qquad (11)
\end{aligned}$$

Similarly, using the quantities $F'$, $\bar{F}'$, $G'$ and $\bar{G}'$ defined in Fig.1, we obtain

$$\begin{aligned}
aT &= \frac{1}{\sqrt{2}\cos\phi}[F'+\bar{F}'+(G'-\bar{G}')\cot\alpha], \\
P_{00} &= \frac{1}{\sqrt{2}\sin\alpha}\sqrt{(F'-\bar{F}')^2+(G'-\bar{G}')^2}, \\
\tan(\delta_{00}-\phi) &= \frac{\bar{F}'-F'}{\bar{G}'-G'}. \qquad (12)
\end{aligned}$$

And in the same manner

$$\begin{aligned}
P_{+-} &= \frac{1}{\sqrt{2}\sin\alpha}\sqrt{(F''-\bar{F}'')^2+\bar{G}''^2}, \\
\tan(\delta_{+-}-\phi) &= -\frac{\bar{F}''-F''}{\bar{G}''}. \qquad (13)
\end{aligned}$$

The assumption that all strong phases are equal implies, $\tan(\delta_{+-} - \phi) = \tan(\delta_{00} - \phi) = \tan(\delta_{+0} - \phi)$. From the above equations, we see that this can be tested. It is also interesting to note that all details about the decay amplitudes can be determined experimentally, even if the strong phases of the penguin amplitudes are not equal. This will provide much needed information for the study of hadronic matrix elements and strong phases which are not possible to be evaluated theoretically at present. Such analysis should be carried out in future experiments.

In conclusion we have shown that improving HX method with inclusion of electroweak penguin contribution there is still a way to determine the phase $\alpha$ and the penguin amplitudes in $B \to \pi\pi$, provided one makes a weaker assumption that all relative strong phases between tree and penguin amplitudes are equal. This method does not depend on the time-dependent measurements of CP asymmetry but needs only the time-independent measurements of the relevant decay rates in $B \to \pi\pi$ which are measurable at B factories or hadronic machines. We also propose a method to extract detailed informations about $B \to \pi\pi$ decay amplitudes using data from other experiments on the weak phases. This informations can be used to study the hadronic matrix elements and strong phases.

This work was supported in part by the Department of Energy Grant No. DE-FG06-85ER40224. XGH was supported in part by Australian Research Council.

# FIGURES

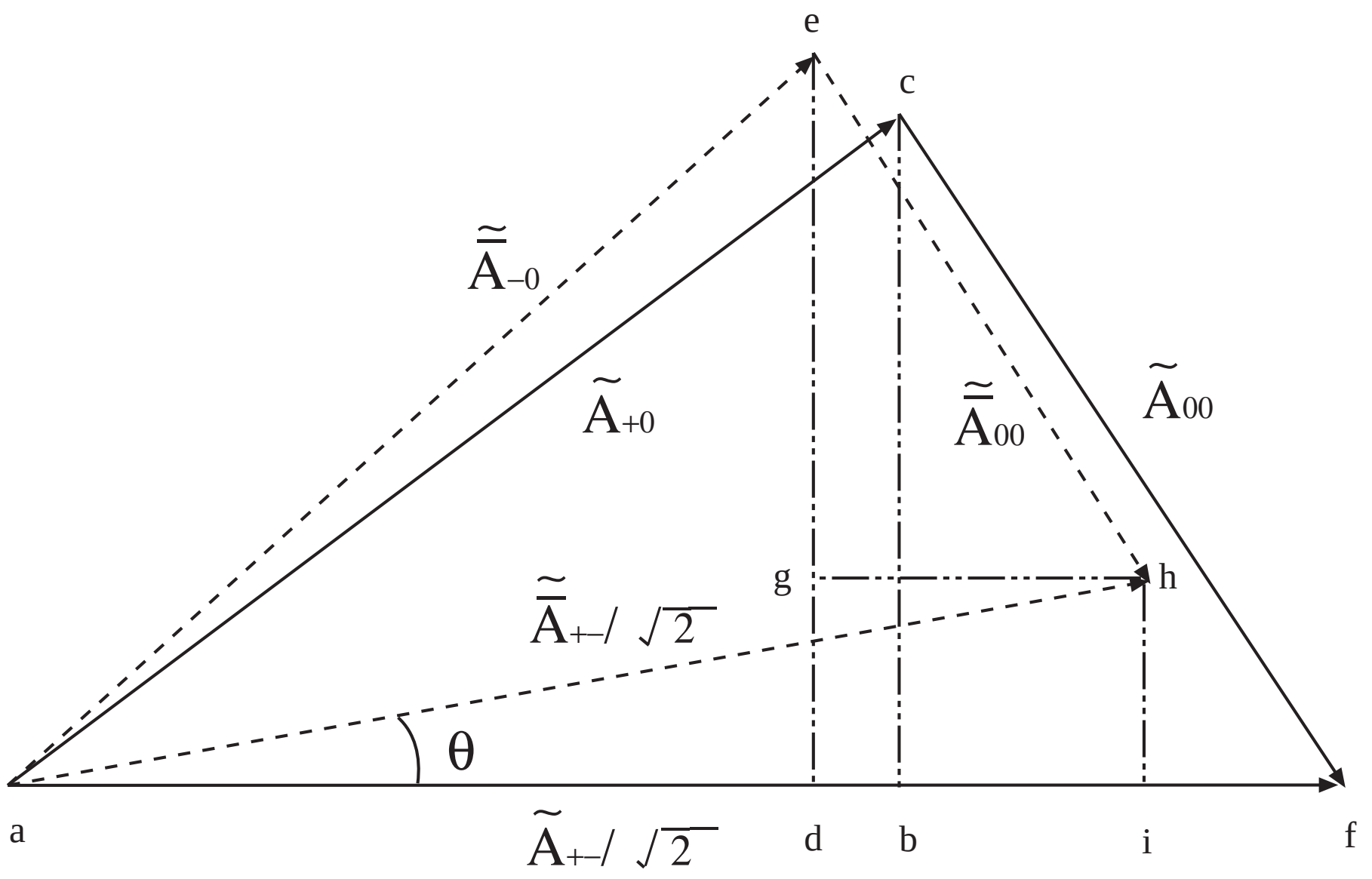

FIG. 1. The two triangles in complex plane constructed by the isospin relations among the rotated amplitudes $\tilde{A}_{ij}$'s and $\tilde{\bar{A}}_{ij}$'s. The quantities $F$, $G$, $\bar{F}$ and $\bar{G}$ are defined as $F = |\overline{ab}|$ (the length between the points $a$ and $b$), $G = |\overline{bc}|$, $\bar{F} = |\overline{ad}|$, and $\bar{G} = |\overline{de}|$, respectively. Similarly, $F'$, $\bar{F}'$, $G'$, $\bar{G}'$, $F''$, $\bar{F}''$, and $\bar{G}''$ are defined as $F' = |\overline{bf}|$, $\bar{F}' = |\overline{gh}|$, $G' = G$, $\bar{G}' = |\overline{ge}|$, $F'' = |\overline{af}|$, $\bar{F}'' = |\overline{ai}|$, and $\bar{G}'' = |\overline{hi}|$, respectively.